# Unravelling a Zigzag Pathway for Hot-Carrier Collection at $CH_3NH_3PbI_3$/Graphene Interfaces

Jin Zhang[1,#,*], Hao Hong[2,#], Jincan Zhang[3,#], Chunchun Wu[2], Hailin Peng[3], Kaihui Liu[2,4,*], and Sheng Meng[1,4,*]

[1]*Beijing National Laboratory for Condensed Matter Physics, and Institute of Physics, Chinese Academy of Sciences, Beijing 100190, P. R. China*

[2]*State Key Laboratory for Mesoscopic Physics, School of Physics, Peking University, Beijing 100871, P. R. China*

[3]*College of Chemistry and Molecular Engineering, Academy for Advanced Interdisciplinary Studies, Peking University, Beijing 100871, China*

[4]*Collaborative Innovation Center of Quantum Matter, Beijing 100190, P. R. China*

\# These authors contributed equally to this work

\* Correspondence: smeng@iphy.ac.cn; khliu@pku.edu.cn; jin.zhang@mpsd.mpg.de

# Abstract

The capture of photoexcited deep-band hot carriers, excited by photons with energies far above the bandgap, is of significant importance for photovoltaic and photoelectronic applications since it is directly related to the quantum efficiency of photon-to-electron conversion. By employing time-resolved photoluminescence and state-of-the-art time-domain density functional theory, we reveal that photoexcited hot carriers in organic-inorganic hybrid perovskites prefer a zigzag interfacial charge-transfer pathway, *i.e.*, the hot carriers transfer back and forth between $CH_3NH_3PbI_3$ and graphene, before they reach a charge separated state. Driven by quantum coherence and interlayer vibrational modes, this pathway at the semiconductor-graphene interface takes about 400 femtoseconds, much faster than the relaxation process within $CH_3NH_3PbI_3$ (in several picoseconds). We further demonstrate that the transfer rate of the pathway can be further enhanced by interfacial defects. Our work provides a new insight for the fundamental understanding and precise manipulation of hot-carrier dynamics at the complex semiconductor-graphene interfaces, paving the way for highly efficient photovoltaic and photoelectric device optimization.

# Introduction

In conventional solar cells, hot charge carriers are generated by absorbing photons with energies larger than the bandgaps, which are generally followed by a quick cooling process, *i.e.*, the hot electrons relax to the band edges of semiconductors.[1-3] If the hot carriers can be captured within a time significantly shorter than the competing energy relaxation and loss processes, light-to-electricity conversion efficiency could be improved effectively. Van der Waals (vdW) heterostructures provide a potential platform for investigating new physics and optoelectrical applications in hot-carrier solar cells.[4-10] To achieve desirable performance in hot-carrier devices based on vdW heterostructures, it is of the central importance to understand and manipulate photoexcited hot carrier dynamics on the ultrafast timescale at interfaces.[11-16]

Benefitting from the excellent optical properties, such as near-infrared to visible tunable bandgap, ultrahigh absorption, near-unity photoluminescence (PL) quantum yield and nanosecond-scale long-lived photocarrier lifetime, organic-inorganic hybrid perovskites, *e.g.,* $CH_3NH_3PbI_3$ (referred to as $MAPbI_3$ hereafter), have emerged as an ideal photoactive material and been successfully used in the light detection and emitter recently.[17-25] With two-dimensional (2D) graphene covered as an electrode, the perovskite-graphene heterostructures have gained substantial interest.[26-29] Several experimental works have reported interfacial carrier dynamics and high photoresponsivity on perovskite-graphene heterostructures.[27-29] Lee and coauthors demonstrated an intriguing photodetector consisting of $MAPbI_3$ and graphene, achieving a relatively high photoresponsivity and an effective quantum efficiency ($5\times10^4$ % at an illumination power of 1 μW).[26] More directly, Hong *et al*. reported that graphene is an excellent carrier acceptor which extracts photoexcited hot carriers of MAPbI$_3$ crystal with interfacial deep-band charge transfer in less than 50 fs, significantly faster than hot-carrier relaxation or cooling process.[27] However, a comprehensive understanding of the dynamic process of photoexcited hot carriers at the semiconductor-graphene interfaces is still rare and at its early stage.

In this article, we utilize tunable two-color pump−probe spectroscopy, time-resolved PL spectroscopy, and the state-of-the-art time-domain density functional theory (DFT) to study the interlayer charge dynamics in model CH$_3$NH$_3$PbI$_3$/graphene (*i.e.*, MAPbI$_3$/graphene) heterostructure. Our results reveal that the ultrafast interlayer charge transfer in the heterostructure takes place on an order of ~400 fs, in good agreement with experimental observations. Surprisingly, we find that photoexcited hot carriers in organic-inorganic hybrid perovskites prefer to undergo a zigzag pathway, *i.e.*, the hot electrons travel back and forth between MAPbI$_3$ and graphene. First-principles simulations yield a thorough understanding of hot-carrier dynamics at a model semiconductor-semimetal interface, which validates our experimental observations. This work not only provides a new and complete physical picture into the hot-carrier collection in MAPbI$_3$/graphene heterostructures on the atomic level, but also sheds light on optimizing their performances towards future applications in high-speed photodetection and high-efficient light-harvesting materials.

**Results and Discussion**

**Experimental ultrafast carrier dynamics at MAPbI$_3$/graphene interfaces**

In our experiments, chemical vapor deposition (CVD) grown graphene monolayer was cleanly transferred on a transmission electron microscopy (TEM) grid without polymer assisted.[30] Then the clean interface of MAPbI$_3$/graphene was fabricated by directly growing single-crystal MAPbI$_3$ crystals on suspending graphene, as shown in Figure 1a. MAPbI$_3$ crystal with a uniform thickness resides on the suspended graphene. Figure 1b depicts the aberration-corrected atomic-resolved TEM image of MAPbI$_3$, further proving a high crystallinity of our samples. No noticeable contaminants are observed around the MAPbI$_3$/graphene region, indicating the high cleanliness of the interface we fabricated. This clean interface promises strong interlayer coupling and an efficient interlayer charge transfer process as discussed below.

Comparing with pristine one, the PL intensity of MAPbI$_3$ is strongly quenched by nearly two orders of magnitude with graphene underneath (Figure 1c). In addition, the carrier lifetime reduces significantly (Figure 1d) on account of the new pathway of ultrafast carrier transfer at the interfaces. From our time-resolved PL experiments, pristine MAPbI$_3$ has a lifetime of ∼25 ns, while the lifetime of MAPbI$_3$/graphene heterostructure is reduced significantly to ∼1.3 ns.

The carrier collection efficiency at the interface can be estimated from either the PL quench ($\eta_{PL} = 1 - \frac{I_{w\,Gr}}{I_{w/o\,Gr}} = 99\%$, where $I_{w\,Gr}$ and $I_{w/o\,Gr}$ indicate the PL intensities of MAPbI$_3$/graphene and pristine MAPbI$_3$, respectively) or the lifetime decreases ($\eta_{PL} = 1 - \frac{\tau_{w\,Gr}}{\tau_{w/o\,Gr}} = 95\%$, where $\tau_{w\,Gr}$ and $\tau_{w/o\,Gr}$ indicate lifetimes of MAPbI$_3$/graphene and pristine MAPbI$_3$, respectively). These two different methods give two different values, with the former one a little larger than the latter one. For the time-resolved PL experiment, the lifetime of cooling carriers at the band edge is measured. It is a little complicated for the case of PL experiments, where deep-band carriers are excited and may directly transfer to graphene right after excitation with high-energy photons. After high energy excitation in perovskite, the carrier relaxation or cooling process inside perovskite is very slow, with a typical time scale of ~50 ps, as shown in the rise curve of Figure 1e, mainly because of phonon bottleneck effects.

By contrast, hot carrier relaxation in graphene (transient absorption signal with probe wavelength at 680 nm) is relatively faster, with a lifetime of 2.1 ps (Figure 1f). Therefore, the dynamic process of the photoexcited hot carriers at the semiconductor-graphene interfaces may be much more complicated and deserves a thorough time-domain atomistic description.

**Potential pathways for hot carrier dynamics at MAPbI$_3$/graphene interfaces**

In Figure 2, we demonstrate three potential pathways involved in the photoinduced hot electron dynamics at the MAPbI$_3$/graphene heterojunction. In the beginning, optical excitation with a high-energy photon ($hv$ > bandgap of MAPbI$_3$) generates a pair of electron ($e^-$) and hole ($h^+$). Then, the hot carrier diffuses to graphene on account of the semiconductor-semimetal contacts. For pathway 1 (Figure 2a), the excited hot electron at MAPbI$_3$ first relaxes to conduction band minimum (CBM) or the band edge of MAPbI$_3$ and then diffuses to graphene due to the band alignment. Pathway 2 is illustrated in Figure 2b, where the photoexcited hot carriers first transfer to CBM of graphene and then relax to the Dirac point. Beyond them, strong interfacial interactions between MAPbI$_3$ and graphene may result in an alternative mechanism (pathway 3 in Figure 2c), where the hot electrons travel forth and back between the semiconductor and graphene in a zigzag pathway. The dominant pathway is determined by comparable time scales of interlayer charge transfer and intralayer hot-carrier relaxation within the two individual components.

Based on our experimental observations as well as results from precious works[26-29], we choose the vertical heterostructure of MAPbI$_3$ and graphene as a model to exploit the interfacial electronic properties and photoexcitation induced carrier dynamics in the semiconductor-semimetal interfaces, see supporting information (SI) for details. More information comes from the projected density of states (PDOS) of the heterostructure (Figure S1). The bandgap of MAPbI$_3$ in the heterostructure calculated with the Perdew-Burke-Ernzerhof (PBE) functional[31] is about 2.2 eV, and graphene is semi-metallic with a zero bandgap. The CBM and valence band maximum (VBM) are both contributed mainly by electronic orbitals from MAPbI$_3$, reflecting a typical semiconductor-semimetal contact. One of the crucial parameters for charge diffusion at the interface is the band offsets at both CBM and VBM states. To validate our results, we compare the PDOS at the PBE level with that based on HSE06 functional (Figure S2).[32] It is observed that PBE functional is accurate enough to describe the spatial distribution of electronic states and the state couplings at MAPbI$_3$/graphene interfaces, which plays a significant role in the dynamic simulations.

**Photoinduced hot carrier dynamics at MAPbI$_3$/graphene interfaces**

In the following, we employ first-principles calculations based on time-domain density functional theory (see methods)[33-41] to explore the photoexcitation induced different processes in MAPbI$_3$/graphene heterostructures. To investigate hot carrier dynamics, we performed non-adiabatic analysis incorporating all orbitals with energies higher than the Fermi level of the heterostructure. Here, two typical hot-electron states are selected: 3.2 eV and 2.9 eV above the Fermi level, respectively, and their dynamics are closely tracked. The hot-electron states are selected according to the pump energy at 410 nm (corresponding to 3.0 eV) used in pump-probe experiments discussed above.

The spatial distributions of the crucial electron states are displayed in Figure 3. It is apparent that photoexcited hot-electron state (Figure 3a) with the energy of about 3.2 eV above the Fermi level of the heterostructure is mainly localized on MAPbI$_3$ crystal, with only a small proportion of the excited state is distributed on graphene. Photoexcited hot-electron state (Figure 3b) is significantly delocalized between the two layers at the time of 23 fs after photoexcitation, attributed to the strong interfacial coupling. The spatial distributions of CBM states localized on MAPbI$_3$ and graphene are also presented in Figure 3. The delocalization of CBM in MAPbI$_3$ is an indication of a relatively strong interfacial coupling.

Pathway 3 can be roughly divided into three sub-steps：photoexcitation induced hot electron travels from perovskite to graphene in an ultrafast time scale (step I). And then, the accumulated carrier at graphene diffuses back to the perovskite (step II). Finally, the electron travels further to graphene with the energy dissipation by the effective electron-phonon and electron-electron scatterings (step III). It should be noted that the hot-carrier relaxations within MAPbI$_3$ and graphene are not included in the separate steps, which tends to contribute to the charge transfer rate considering they are naturally involved in the entangled dynamics.

To gain quantitative information, we interrogate carrier populations on graphene orbitals at different times upon photoexcitation. The evolution of the localization is shown in Figure 4, where the population represents the fraction of the photoexcited charge density on the graphene sheet. At t=0 fs, the two photoexcited states are mostly distributed on MAPbI$_3$, with 10% (for the state of 3.2 eV) and 20% (for the state of 2.9 eV) the excited hot electrons diffuse to graphene in an ultrafast time scale (< 20 fs for both states, Figure 4a). The population on graphene increases from 10% to 34% for the first charge diffusion process. In the following step, the hot electron diffuses back to MAPbI$_3$ and the population on graphene decreases to 23% at 50 fs. This process is not surprising because similar processes were observed that photoexcited hot carrier in graphene can inject into semiconductors (*i.e.*, WS$_2$) in about 25 fs, accompanied with the quantum yield as high as ~50%.[8,42-43] Afterward, the hot electrons diffuse from MAPbI$_3$ to graphene with a large percentage (~86%) within 600 fs. Simultaneously, the energy of the state decreases continuously from 2.9 eV to 0 eV upon photoexcitation (Figure 4b). Consequently, the time scale of the whole process is 400 fs with exponential fitting.

In contrast, the charge population on graphene increases significantly in 23 fs (from 20% to 48%) for the excited state with the energy of 3.2 eV above the Fermi level. Then, this value drops to 30% at t=75 fs. The total time scale of the photoexcited process is observed to be about 400 fs, accompanied by the energy decreasing continuously from 3.2 eV to 0 eV. These findings reveal there exists a new robust relaxation process (pathway 3) on a time scale of ~400 fs, validating the experimentally observed ultrafast broadband charge transfer from MAPbI$_3$ to graphene.[27]

To have a clear understanding in the mechanism of the entangled processes, we separate the distinct processes by involving only relative states in the vicinity of the excited electron state. In other words, the states with energies far away from the selected excited states are switched off in the real-time charge propagation. For step I (Figure 4c), the time scale of charge diffusion is about 23 fs from MAPbI$_3$ to graphene, in good agreement with experimental observation (< 50 fs). The time evolution data are fitted with exponential equation $\chi = a + b * exp\,(-t/\tau)$, where $\tau$ is the lifetime and $\chi$ is the population of the carriers, namely, the integral of excited electron density on graphene orbitals. Then in the following 30 fs, the electron transfer process from graphene back to MAPbI$_3$ takes place (step II, as shown in Figure 4d).

On the other hand, there is an alternative cooling routine for the hot carriers: rapid electron-hole annihilation inside the semimetallic graphene with the assistance of effective electron-phonon and electron-electron scattering. This energy loss competes with charge separation and the dominant one determined by relative time scales. It is obtained the relaxation timescale is 800 fs for the hot electron with the energy of ~3.0 eV above the Fermi level in monolayer graphene (Figure S3), significantly slower than the interfacial charge transfer process (30 fs). The lifetime is on the same order of magnitude with previous experimental studies (~50 fs),[27] accounting for the strong interfacial interactions between MAPbI$_3$ and graphene and the assistance of interlayer phonon vibrations. For MAPbI$_3$, it is shown that the cooling process of hot carriers apparently takes a longer time (~50 ps). Thus, the energy relaxation within both graphene and MAPbI$_3$ are much slower than the interface-assisted carrier dynamics, facilitating the novel pathway where hot electrons in the heterostructure would travel back and forth.

In order to accurately account for the band-edge carrier transfer, Figure 4e illustrates the excited electron at the CBM of MAPbI$_3$ further diffuses to graphene in 92 fs, in excellent accordance with the previous experimental result (110 fs)[27]. It should be noted there is a slight discrepancy between the total time scale (~170 fs) and the summation (~150 fs) of the three separate steps. This difference is attributed to the relaxation process in MAPbI$_3$ and graphene, which is not taken into account in the three separate steps. These findings further rationalize the distinct charge dynamics for the band-edge electron (with the pump energy at 1.5 eV or the wavelength of 820 nm) and hot electron (pump energy of ~3.0 eV, 410 nm) from MAPbI$_3$ to graphene using PL excitation spectrum in experiments. Therefore, the results validate the novel and robust pathway 3, instead of the conventional decaying pathways 1 and 2 at the vdW interfaces.

**MAPbI$_3$/graphene interfaces with defects**

Carrier transfer rate is determined by the interactions between MAPbI$_3$ and graphene surfaces. The strength of interfacial interaction is influenced by MAPbI$_3$/graphene geometry and the spatial separation. Challenges retain in real optoelectronic and photovoltaic applications because interfacial defects are unavoidable in MAPbI$_3$/graphene heterostructures. Wang and coauthors reported that MAPbI$_3$ could be either *n*- or *p*-doped by changing the ratio of methylammonium halide and lead iodine which are the two precursors for perovskite formation.[28] However, the underlying response of the zigzag charge-transfer process discussed above with respect to defects of this intriguing system is still lacking.

In this regard, we take MA defect as an example to demonstrate its effect on the peculiar pathway for hot-carrier collection at MAPbI$_3$/graphene interfaces. Defects are generated at the interfaces with a concentration of 10%. For the heterostructure with MA defects, the electron acceptor state is entirely significantly localized on the graphene layer. Both types of vacancies perturb the symmetry of the heterostructure and give rise to phonon modes that are not available in the pristine system. Figure S3 displays PDOS and photoinduced processes for perovskite-graphene heterostructures with interfacial MA defects. It is observed iodine vacancies are inclined to lower the Fermi level of MAPbI$_3$, resulting in a smaller band offset (0.5 eV) between MAPbI$_3$ and graphene. In contrast, we see MA defects amplify the band offset to 1.5 eV, indicating a large driving force for interfacial charge transfer.

We also demonstrate the charge population of the excited hot electron at MAPbI$_3$ with MA defects (initial photoexcited energy of 2.4 eV above the Fermi level) and the corresponding energy evolution of the excited hot electrons. Our results demonstrate that the ultrafast interlayer charge transfer takes place with the transfer time on an order of ~150 fs. This indicates interfacial MA defects can significantly enhance the transfer rate in the zigzag pathway, which is attributed to the assistance of interfacial phonon modes. The findings thereby allow us to understand interfacial charge dynamics at the perovskite-graphene heterostructures at the atomic level.

**Electron-phonon interactions in charge dynamics**

To further understand the peculiar pathway for photoexcited hot-electron dynamics, more information comes from the effect of decoherence and phonons are analyzed in Figure 5. The decoherence effects are particularly important because the loss of coherence occurs generally much faster than the corresponding quantum transition. The time scales of coherence loss or pure dephasing determine the homogeneous linewidths of the corresponding optical transitions. For the hot electrons in the heterostructure, the dephasing takes place in about 5.1 fs, indicating electron-phonon interaction plays a vital role in the overall hot-electron dynamics.

Furthermore, the Fourier transformation of the energy difference along the molecular dynamics would permit a quantitative analysis of the crucial phonon vibrations mediated in the process. We obtain that the Pb-I bond stretching mode (~90 cm$^{-1}$) and the liberation of the organic cations (CH$_3$NH$_3{}^+$) at ~120 cm$^{-1}$ in perovskite-graphene heterostructure indeed plays a dominant role in ultrafast hot-electron dynamics.[44-45] In addition, the frequency around 1500 cm$^{-1}$ is attributed to carbon-carbon stretching modes in graphene. Therefore, phonon modes in both MAPbI$_3$ and graphene are involved to assist the emergent zigzag-like hot-electron transfer route. This observation, together with strong coupling between MAPbI$_3$ and graphene layers, elucidates the mechanism of the newly-discovered hot–electron collection channel.

**Discussion**

Our joint experimental and theoretical study presented in this work has broad implications in the semiconductor-graphene heterostructures for the future hot-carrier solar cells. In photoelectric devices, hot carriers are generated by absorbing the photons with energies larger than the corresponding bandgap of the semiconductor. The efficiency of devices is greatly affected by the fast cooling process because it limits the charge collection efficiency of hot carriers. Effective interfacial electronic couplings and the collective phonon modes result in a novel zigzag pathway, which is on the order of the magnitude faster than the electron cooling process within the semiconductors. To apply the graphene as a 2D electrode, the zigzag pathway for hot carriers may lead to additional challenges: the injected hot carriers on graphene are able to transfer back to the semiconductors, reducing the extraction efficiency of hot carriers into external circuits. The challenge can be overcome by adhering to another electrode to extrude the hot carriers from graphene, with a more competitive charge collection efficiency.

It should be noted that the theoretical analysis is in good consistence with the coherent hot carrier dynamics demonstrated by the ultrafast broadband charge collection achieved simultaneously at the clean graphene/organic-inorganic halide perovskite interfaces. We should also point out that some intriguing carrier-relaxation processes for hot carriers have been reported in some recent studies involving molecule-semiconductor interfaces.[46] It is also demonstrated that the heterostructures can mediate the charge dynamics at 2D heterostructures effectively, leading to a novel pathway for hot carrier relaxation through interlayer hopping.[47] However, it is the first time to demonstrate the interfacial multiple-step carrier dynamics modulated by interlayer phonon modes and electronic state couplings.

Beyond the zigzag pathway, we envision the photoexcited hot carriers with a higher energy at semiconductor-graphene interfaces may experience even more complicated pathways (*e.g.*, incorporating more and delicate interfacial charge transfer processes). Thus, to fully understand the fate of photoexcited carriers, one should consider ultrafast intralayer/interlayer carrier scatterings, interfacial excitons and charge transfer processes at the semiconductor-graphene heterostructures.

In summary, we demonstrate that photoexcited hot carrier dynamics in MAPbI$_3$/graphene heterostructures using ultrafast transient absorption spectrum, time-resolves PL and *ab initio* time-domain DFT simulations. We reveal the strong interfacial interactions between MAPbI$_3$ and graphene result in a novel zigzag hot-carrier cooling pathway, where the hot electron travels back and forth between semiconductors and graphene. Interfacial MA defects can significantly enhance the transfer rate of the zigzag pathway, which is attributed to the assistance of interfacial phonon modes. Our calculations not only explain the experimentally observed ultrafast carrier dynamics upon photoexcitation at both band edges and deep bands, but also pave a new way to manipulate the hot carriers in semiconductor-graphene heterostructures in optoelectronic and photovoltaic applications. Therefore, improved device performance of hot-carrier solar cells may be achieved given that the photoexcited carriers in graphene can be extracted into electric circuits to minimize the loss in carrier conductors.

# Methods

**Sample preparation.** Single-crystal graphene was grown on Cu foil (Alfa-Aesar, No. 46365) by chemical vapor deposition method. For $CH_3NH_3PbI_3$ synthesis, $CH_3NH_3I$ and $PbI_2$ were dissolved in isopropanol at a 3:1 molar ratio with final concentrations of ~40 %. Then, the graphene grid was floated on the surface of perovskite solution with the graphene face contacting the solution instead of rinsing it in the solution. This procedure guarantees the ultra-clean $MAPbI_3$/graphene interface.

**Transient absorption and time-resolves PL experiments.** The pump-probe measurements were performed by femtosecond pulses (~120 fs, 80 MHz) generated by a Ti:sapphire oscillator (Spectra-Physics Mai Tai laser) and an optical parametric amplifier (OPO). In the measurement of transient absorption of graphene, we pumped at 410 nm and probed at 680 nm. Those two pulses are separated in the time-domain by a controllable delay-time and focused onto the sample by a 100 X/0.9 NA objective. After collection of the reflected pulses, 460 nm long-pass filter was used to filter out the pump pulse. The transient absorption signal was recorded by a PMT and a lock-in amplifier with reflective geometry. The time-resolved PL was excited by pulses from Coherent Laser setup (250 KHz, 70 fs) at 400 nm. PL signal of target wavelength was selected by a 460 long-pass filter and a spectrometer (with resolution of $\pm2$ nm) after photoexcitation and collection. Then, the time-resolved PL signal was acquired by a single-photon APD (PicoQuant Company, TDA 200) combining with a TCSPC module (TimeHarp 260 PICO Single)[27].

**Treating ground-state properties**. The calculations were performed using density functional theory as implemented in the Vienna *ab initio* simulation package[31-34] using a projector-augmented wave (PAW) pseudopotential in conjunction with the Perdew-Burke-Ernzerhof (PBE) generalized gradient approximation for the exchange-correlation energy and van der Waals density functional in the form of opt88[35-36]. Both lattice constants and atomic positions were relaxed with an energy cutoff of 400 eV until all residual forces are less than $10^{-2}$ eV/Å and the total energy variation is less than $10^{-4}$ eV (the vacuum layer is set to be 15 Å). To perform molecular dynamics calculations, the lattice parameters of the pristine heterostructure are set to 12.70 Å and 24.70 Å for the two in-plane directions with 168 atoms for $MAPbI_3$ and 120 carbon atoms for graphene, resulting in a negligible strain at the interface. After full relaxation, there is an obvious bending in the Pb-I frame toward graphene, which is ascribed to the strong interfacial interaction between $MAPbI_3$ and graphene.

**Treating nonadiabatic molecular dynamics.** To perform time-domain DFT calculations[37-44], the structure optimization, and molecular dynamics (MD) were performed using only the Γ-point of the first Brillouin zone since a large supercell was used. After geometry optimization at 0 K, the $MAPbI_3$/graphene heterostructures with and without defects were heated to 300 K, corresponding to the temperature used in the experiments. After that, 3 ps long adiabatic MD trajectories were generated in the microcanonical ensemble with a 1 fs timestep. To simulate the electron/hole dynamics, 1000 geometries were selected randomly from each adiabatic MD trajectory. They were used as initial conditions for calculating nonadiabatic couplings, which was performed using fewest-switches surface hopping in the classical path approximation and with the decoherence correction. The atomic timestep was set to 1.0 fs and the electronic time step was 1.0 attosecond. See supplementary materials for more details.

**Acknowledgements**

This work was supported by National Key Research and Development Program of China (Grant Nos. 2016YFA0300902, 2015CB921001 and 2016YFA0300903), National Natural Science Foundation of China (Grant Nos. 11774396, 11474328 and 51522201), and "Strategic Priority Research Program (B)" of Chinese Academy of Sciences (Grant No. XDB07030100).

**Author contributions**

S.M. and K.L. designed the research. Most of the calculations and analysis were performed by Jin Zhang with contributions from all authors. H.H., Jincan Zhang and C.W. performed the experiments under supervision of K.L. and H.P. All authors contributed to the analysis and discussion of the data and the writing of the manuscript.

*Conflict of Interest:* The authors declare no competing financial interest.

# Figures and Captions

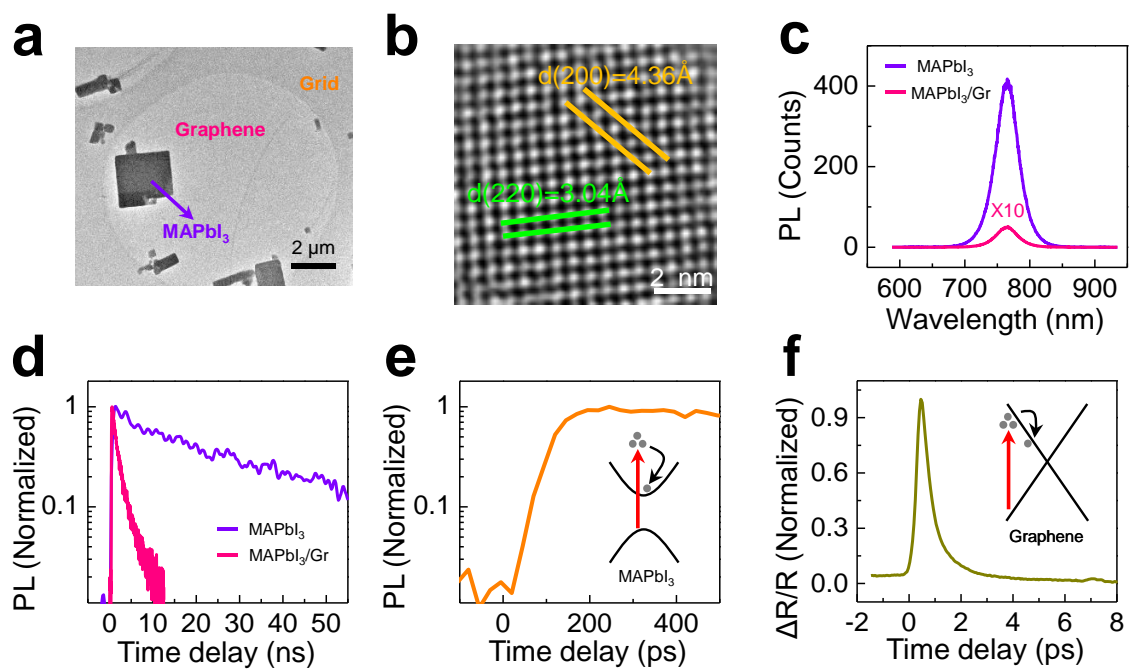

**Figure 1 | Experimental samples and pump-probe measurements of MAPbI$_3$/graphene heterostructure.** **(a)** Transmission electron microscopy (TEM) image of the suspended MAPbI$_3$/graphene. The scale bar is 2 μm. **(b)** Representative aberration corrected TEM image of the lattice structure of MAPbI$_3$. Typical lattice parameters are illustrated in the figure. **(c)** Photoluminescence (PL) spectra of MAPbI$_3$ with (MAPbI$_3$/Gr) and without graphene (MAPbI$_3$). **(d)** Time-resolved PL of pristine MAPbI$_3$, MAPbI$_3$ grown on polymer-contaminated graphene, respectively. **(e)** Time-resolved PL of pristine graphene and MAPbI$_3$ grown on clean graphene. **(f)** Evolution of transient absorption signal with probe wavelength at 680 nm. The selected trace undergoes biexponential decay with lifetimes of 0.3 ps and 2.1 ps, representative dynamics from graphene.

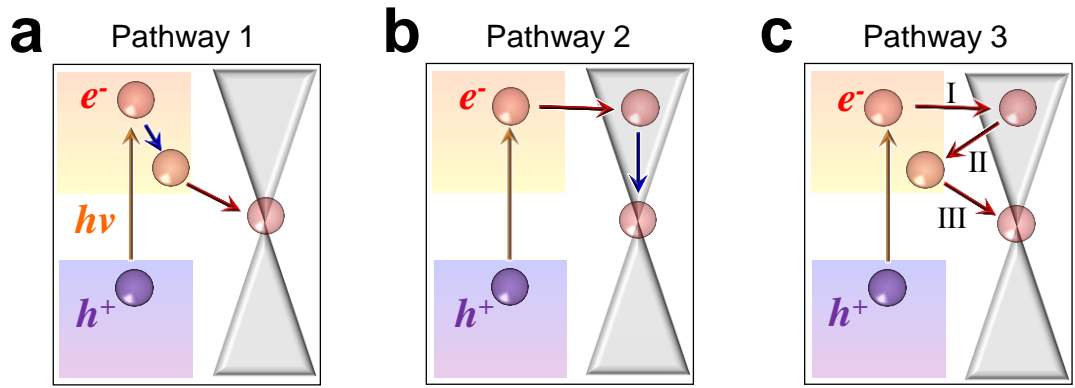

**Figure 2 | Schematic illustration of the photoinduced hot carrier dynamics at MAPbI$_3$/graphene heterostructure. (a)** Pathway 1: photoexcited hot electron ($e^-$) in MAPbI$_3$ first relaxes to the band edge and then transfer to graphene. **(b)** Pathway 2: photoexcited hot electron in MAPbI$_3$ first transfers to graphene and then relaxes among graphene. **(c)** Pathway 3: photoexcited hot electro in MAPbI$_3$ first transfers to graphene across a zigzag pathway at semiconductor-graphene interfaces. I, II and III indicate the separate steps of the pathway. See the main text for details.

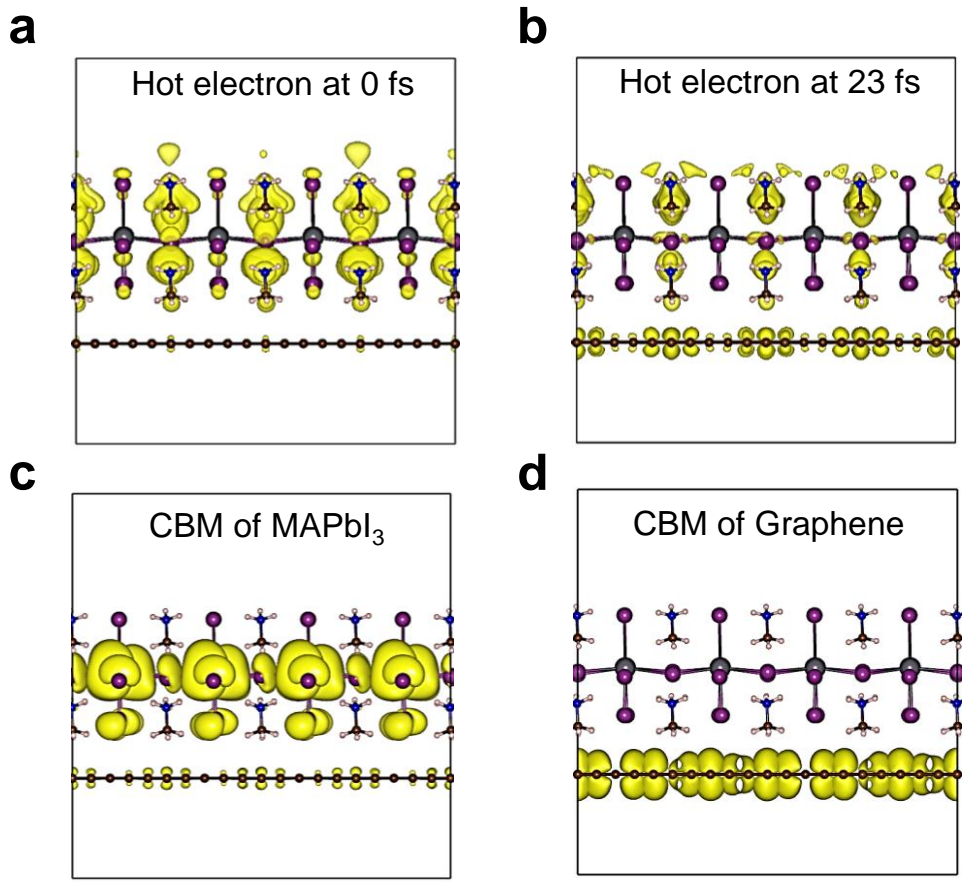

**Figure 3 | Charge distributions of the donor and acceptor states for the electron dynamics in MAPbI$_3$/graphene heterostructure. (a)** The hot-electron state with the energy of 3.2 eV above the CBM after photoexcitation at 0 fs. **(b)** Photoexcited hot-electron state transferred from perovskite to the graphene at 23 fs. The state is significantly delocalized at both MAPbI$_3$ and graphene, showing an intermedia state during charge transfer. **(c, d)** Transferred electron states at the CBM of perovskite and graphene, respectively. The charge distributions are at a contour level of $7 \times 10^{-4}$ e/Å$^3$. Large black (purple) spheres represent Pb (I) atoms while the small brown, blue and pink spheres represent carbon, nitrogen, and hydrogen atoms, respectively.

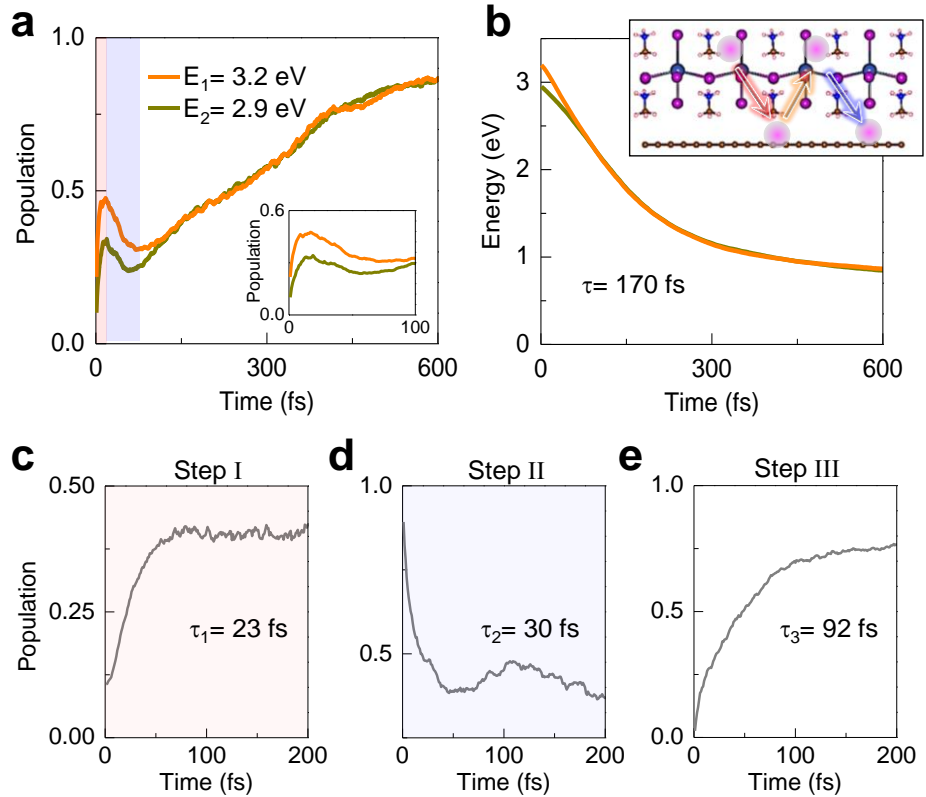

**Figure 4 | Hot carrier dynamics at MAPbI$_3$/graphene interfaces.** (**a**) The population at MAPbI$_3$ of excited hot electron with initial photoexcited energies of 3.2 eV and 2.9 eV above the Fermi energy as a function of time upon optical excitation. For clarity, the inset shows the zoom-in region from 0 to 100 fs. (**b**) The corresponding energy evolution of excited hot electrons. For both photoexcited states, the energies decays to lower-energy states at timescales of 170 fs. The inset cartoon exhibits the zigzag pathway in real space. (**c**) Evolution of population of the hot electron from MAPbI$_3$ to graphene in 23 fs (step I) for the initial state with the energy of 3.2 eV. The trends for the two selected hot-electron states are similar. (**d**) Evolution of population of the transferred hot-electron state on graphene diffuses to MAPbI$_3$ in an ultrafast time scale of 30 fs (step II). (**e**) the excited electron at CBM transfers to graphene in 92 fs (step III). The dots in different colors are from our simulations while the time scales are fitted with exponential functions. Inset shows the schematic of the photoinduced carrier dynamics of MAPbI$_3$/graphene heterostructure in real space.

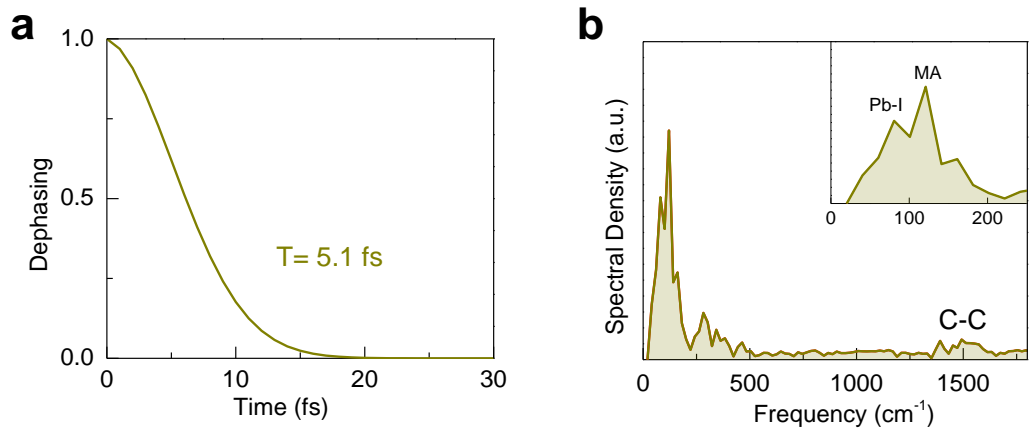

**Figure 5 | Phonon vibration facilitated hot-electron transfer at MAPbI$_3$/graphene interfaces.** (**a**) Pure-dephasing function of hot-electron dynamics. (**b**) Spectral density of the phonon modes evolved in the dynamics: Pb-I bending, MA molecule rotation and carbon-carbon stretching modes are crucial in the carrier dynamics. The inset presents the key phonon branches of MAPbI$_3$ involved in the charge dynamics.

# Table of contents

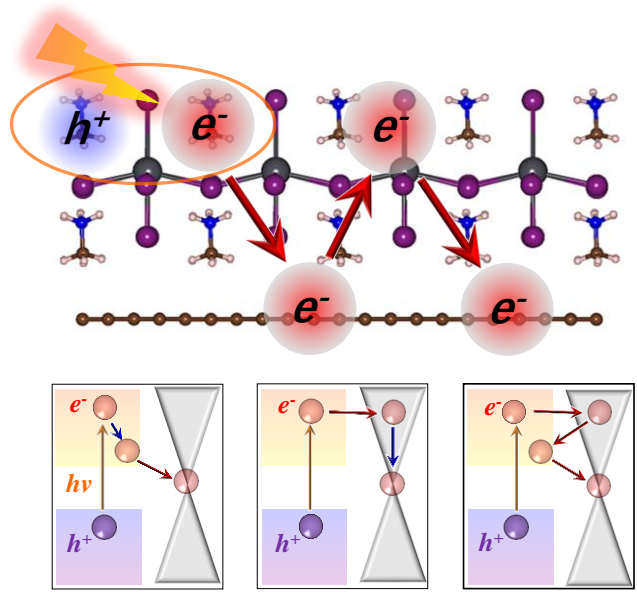